\documentclass[conference]{IEEEtran}
\IEEEoverridecommandlockouts

\usepackage{cite}
\usepackage{amsmath,amssymb,amsfonts}
\usepackage{algorithmic}
\usepackage{graphicx}
\usepackage{textcomp}
\usepackage{xcolor}

\usepackage{svg}
\usepackage[hidelinks]{hyperref}
\usepackage{tikz}
\usepackage{booktabs}
\usepackage{listings}
\usepackage[ruled,linesnumbered,longend]{algorithm2e}
\usepackage{fancyvrb}
\usepackage{fvextra}
\usepackage{enumitem}
\usepackage[nameinlink]{cleveref}
\usepackage{censor}

\definecolor{codegreen}{rgb}{0,0.6,0}
\definecolor{codegray}{rgb}{0.5,0.5,0.5}
\definecolor{backcolour}{HTML}{FAFAFA}
\definecolor{comment}{HTML}{90A4AE}
\definecolor{keyword}{HTML}{E53935}
\definecolor{string}{HTML}{91B859}

\newcommand{\eval}{\textit{Ecoscape}}

\makeatletter
\newcommand{\newlineauthors}{%
  \end{@IEEEauthorhalign}\hfill\mbox{}\par
  \mbox{}\hfill\begin{@IEEEauthorhalign}
}
\makeatother

\lstdefinestyle{code}{
    commentstyle=\color{comment},
    keywordstyle=\color{keyword},
    numberstyle=\tiny\color{codegray},
    stringstyle=\color{string},
    basicstyle=\ttfamily\footnotesize,
    breakatwhitespace=false,         
    breaklines=true,                 
    captionpos=b,                    
    keepspaces=true,                 
    showspaces=false,                
    showstringspaces=true,
    showtabs=false,                  
    tabsize=2
}

\lstdefinelanguage{Yaml}{
    morecomment=[l]{\#},
    moredelim=**[il][\color{black}{:}]{:}
}

\def\BibTeX{{\rm B\kern-.05em{\sc i\kern-.025em b}\kern-.08em
    T\kern-.1667em\lower.7ex\hbox{E}\kern-.125emX}}
\begin{document}

\title{Ecoscape: Fault Tolerance Benchmark for Adaptive Remediation Strategies in Real-Time Edge ML 
}

\author{
\IEEEauthorblockN{Hendrik Reiter}
\IEEEauthorblockA{\textit{AG Software-Engineering} \\
\textit{Christian-Albrechts-University}\\
Kiel, Germany \\
0009-0003-8544-0012}
\and
\IEEEauthorblockN{Ahmad Rzgar Hamid}
\IEEEauthorblockA{\textit{Maersk Mc-Kinney Moller Institute} \\
\textit{University of Southern Denmark}\\
Odense, Denmark \\
0000-0002-1768-2453}
\and
\IEEEauthorblockN{Florian Schlösser}
\IEEEauthorblockA{\textit{AG Software-Engineering} \\
\textit{Christian-Albrechts-University}\\
Kiel, Germany \\
stu240349@mail.uni-kiel.de}
\newlineauthors
\IEEEauthorblockN{Mikkel Baun Kjærgaard}
\IEEEauthorblockA{\textit{Maersk Mc-Kinney Moller Institute} \\
\textit{University of Southern Denmark}\\
Odense, Denmark \\
0000-0001-5124-744X}
\and
\IEEEauthorblockN{Wilhelm Hasselbring}
\IEEEauthorblockA{\textit{AG Software-Engineering} \\
\textit{Christian-Albrechts-University}\\
Kiel, Germany \\
0000-0001-6625-4335}
}

\maketitle

\begin{abstract}
Edge computing offers significant advantages for real-time data processing tasks, such as object recognition, by reducing network latency and bandwidth usage.
However, edge environments are susceptible to various types of fault. A \textit{remediator} is an automated software component designed to adjust the configuration parameters of a software service dynamically. Its primary function is to maintain the service's operational state within predefined Service Level Objectives by applying corrective actions in response to deviations from these objectives. Remediators can be implemented based on the Kubernetes container orchestration tool by implementing remediation strategies such as rescheduling or adjusting application parameters. However, currently, there is no method to compare these remediation strategies fairly. 

This paper introduces \eval{}, a comprehensive benchmark designed to evaluate the performance of remediation strategies in fault-prone environments. Using Chaos Engineering techniques, \eval{} simulates realistic fault scenarios and provides a quantifiable score to assess the efficacy of different remediation approaches.
In addition, it is configurable to support domain-specific Service Level Objectives. We demonstrate the capabilities of \eval{} in edge machine learning inference, offering a clear framework to optimize fault tolerance in these systems without needing a physical edge testbed.

\end{abstract}


\section{Introduction}
\label{sec:Introduction}

Edge computing has emerged as a transformative computing paradigm facilitating data processing proximal to the source. This approach yields significant advantages~\cite{Chen2020}, including reduced latency, bandwidth consumption, and improved data privacy~\cite{Meuser2024}. The Edge-Cloud Continuum is a computing paradigm integrating edge and cloud resources into a unified, heterogeneous infrastructure. It enables efficient distribution of workloads by leveraging low-latency, localised processing of edge computing alongside the scalability and computational capacity of the cloud~\cite{Meuser2024}. The continuum introduces resource-aware execution across diverse application scenarios.

Machine Learning-based (ML) data processing pipelines performing tasks such as computer vision, natural language processing, or analyzing sensor data may benefit from the reduced latency offered by edge computing. This paradigm is particularly advantageous for real-time applications, where latency requirements are paramount~\cite{Bian2022}. ML algorithms usually discuss the trade-off between machine learning model efficiency and accuracy~\cite{Bian2022}. However, techniques such as pruning~\cite{Molchanov2016}, lower bit precision~\cite{Jain2018}, or early exit~\cite{Panda16} support real-time requirements.

Service Level Objectives (SLOs) define performance targets for software systems and services. They also decide the appropriateness of responses to system faults, which requires determining optimal decision-making strategies.
Assume a cluster of nodes, each being an edge or a cloud node. Edge nodes possess limited resources but are located closer to the data source. Cloud nodes are significantly more resourceful but located further away. If a network congestion fault happens, the communication with the cloud will be impeded, resulting in a substantially increased processing latency. In scenarios where accuracy is prioritized, the system might persist in executing computations on the cloud node despite the latency penalty. Conversely, when latency is the primary SLO, the system could dynamically shift processing to a smaller, resource-constrained node closer to the data source, thereby sacrificing some accuracy for improved responsiveness. This perspective is particularly relevant, as faults are ubiquitous in edge computing systems~\cite{Pourreza2023}. These faults manifest in various forms, such as crash, performance, message loss, network partitioning, or byzantine faults. 

The orchestration and management of containerized applications are increasingly based on Kubernetes, a platform that provides automated deployment, scalable resource allocation, service discovery, and fault remediation. Its extensible architecture allows custom implementations of bespoke functionalities, enabling optimization tailored to domain-specific operational requirements. Prominent examples include the development of custom autoscaling mechanisms~\cite{Tran2022} and scheduling algorithms~\cite{Senjab2023}. In particular, custom schedulers~\cite{Rejiba2022} are frequently deployed to enhance the performance of edge computing deployments, addressing critical parameters such as QoS, topological awareness, co-location optimization, data locality awareness, and batch processing efficacy. Furthermore, Kubernetes can respond to fault conditions and restore the system to a functional state. These methods also include the custom autoscalers and schedulers we refer to as remediation strategies in the remainder of the paper. 

Current research lacks comprehensive and standardized methodologies for evaluating remediation strategies within Kubernetes, particularly concerning the adherence to multiple SLOs. This paper introduces \eval{}, a benchmark designed to assess the efficacy of remediation strategies under intentionally injected faults. The name '\eval', which denotes the "organizational structure [...] of an ecosystem"\footnote{\href{https://en.wiktionary.org/wiki/ecoscape}{https://en.wiktionary.org/wiki/ecoscape}}, reflects the heterogeneous and hierarchical layout in edge computing environments. The \eval{} benchmark facilitates the comparative analysis of diverse actions that a Kubernetes remediator can execute to maintain predefined SLOs. These actions encompass scheduling adjustments, scaling operations, modifications to system parameters, and algorithmic implementations, such as using pruned ML models. Previous investigations \cite{Hamid24} have demonstrated that these reconfiguration actions can substantially impact SLO compliance. Furthermore, we present an ML inference use case, specifically object recognition, to validate the remediator's capabilities. \eval{} is tailored for software engineers seeking to evaluate their Kubernetes algorithms within simulated fault-prone edge computing scenarios, thereby obviating the necessity for physical edge testbeds.

The remainder of this paper is structured as follows. \Cref{sec:RelatedWork} introduces related work, and \Cref{sec:Foundation} presents the foundations and proposes the problem statement. \Cref{sec:Ecoscape} demonstrates the design of \eval. \Cref{sec:Evaluation} shows the capabilities of \eval{} according to the object recognition use case. \Cref{sec:Discussion} discusses the design decisions, while \Cref{sec:Conclusion} concludes the paper. 

\section{Related Work}
\label{sec:RelatedWork}

The relevant literature can be categorized into three principal domains: Fault Tolerance benchmarks, Edge Computing benchmarks, and Kubernetes deployment in edge computing environments.

There are several methodologies in the realm of fault tolerance benchmarking. In particular, \textit{Frisbee}~\cite{Nikolaidis2021} presents a tool for automated fault tolerance testing in cloud environments, employing artificial fault injection. Although conceptually aligned with our \eval{} approach, Frisbee diverges in scope by not explicitly addressing edge computing or Kubernetes remediation strategies. In addition, \textit{Frisbee} does not provide quantitative performance metrics. 
Chaos Engineering focusing on Edge computing is achieved in tools such as µChaos~\cite{Kalka2024}. 
However, this tool relies on ZephyrOS, a specific operating system requiring physical hardware for experimental validation. Alternatively, Edge Cloud simulation tools, exemplified by EdgeCloudSim~\cite{Sonmez2017}, offer the capability to emulate the edge-cloud continuum, thus eliminating the reliance on physical infrastructure. Nevertheless, these discrete simulation approaches exhibit limited portability to real-world environments compared to implementations grounded on Kubernetes.

The application of online ML in edge computing encompasses a diverse array of use cases, including, but not limited to, computer vision, real-time speech recognition, and autonomous vehicle operation~\cite{Moreschini2022, Bian2022}. The DeFog~\cite{McChesney2019} benchmark suite presents six representative tasks, such as speech-to-text conversion and real-time face detection on video streams, to elucidate the implications of varying infrastructure configurations, specifically edge-cloud, edge-only, and cloud-only deployment modes. KFIML~\cite{Wan2022}, an application designed for real-time ML at the edge, employs an architecture comprising Kafka brokers and ML tasks executed within the Apache Flink framework. Thus, this setup comes close to the envisioned infrastructure of the \eval{} use-case. The efficacy of this approach is validated through empirical experimentation on a physical testbed.
Furthermore, investigations have explored the optimization of Kubernetes deployments within edge computing environments. Various strategies have been proposed to minimize latency in such deployments. For instance, implementing custom network-aware scheduling techniques within Kubernetes has demonstrated the potential to achieve substantial latency reductions\cite{Santos2019}. Additionally, studies have shown that optimizing Kubernetes configurations within fog computing setups can significantly decrease failover times \cite{Eidenbenz2020}. Determining optimal deployment strategies for edge systems, particularly when confronted with conflicting SLOs as latency and power consumption, has also been a subject of research \cite{Deng2016}.
\section{Foundation \& Problem Statement}
\label{sec:Foundation}

In system reliability, there is a distinction between faults and failures~\cite{Salfner2010}. A failure is a deviation from the delivered service to the intended service. In contrast, a fault represents the underlying problem that can potentially result in a failure. A system is deemed fault-tolerant when faults do not precipitate a failure in service delivery.

Chaos Engineering \cite{Owotogbe2024} has been established as a systematic methodology for evaluating the influence of faults on service behavior through controlled experimentation within production environments. This approach adheres to an iterative process, commencing with defining the system's steady state. Subsequently, a hypothesis concerning the system's behavior under induced chaotic conditions is formulated. Following this, experiments are planned and executed. Finally, any observed anomalies or deviations from the hypothesized behavior are addressed through appropriate remediation strategies.

SLOs~\cite{Qazi2024} serve as a means to define the desired state of a system. This requires the establishment of quantifiable metrics, referred to as service level indicators (SLIs). The SLO integrates the SLI with a threshold value, the transgression of which signifies an SLO violation. A Service Level Agreement (SLA) specifies the temporal duration for which an SLO must be maintained, e.g., ensuring that $99\; \%$ of request latencies remain below 2.5 seconds. Hence, in the occurrence of faults, the SLO may be violated for the time specified by the SLA.
In real-time ML applications at the edge, typical SLOs may include processing latency, classification accuracy, and energy consumption~\cite{Bian2022}.


This paper proposes a benchmark to evaluate the performance of remediators within edge computing environments. The approach is based on the following assumptions regarding the remediator and the edge computing infrastructure: (0) the infrastructure consists of multiple zones with heterogeneous proximal computing nodes; (1) a remediator orchestrates a distributed service across those zones; (2) tasks are portable and can be executed on any node within the system; (3) the load is distributed among the zones; (4) the distributed service is subject to various fault conditions such as tasks exceeding node computational capacity and constrained network communication; and (5) the remediator can dynamically reconfigure the distributed system. 

A benchmark~\cite{Hasselbring2021} is defined as a methodology designed to investigate the quality of service of a software system under a specified workload. This investigation is conducted in an automated and reproducible manner within a precisely described execution environment. Crucially, the benchmark must articulate measurable quality attributes that enable quantitative evaluation. A comprehensive replication package, including the workload, system configuration parameters, and specifications for experiment repetitions or execution duration, is essential to ensure reproducibility. Ideally, the benchmark should also provide diverse datasets, analysis scripts, and extensive documentation to facilitate broader applicability and interpretation. Furthermore, a large, independent, open-source community dedicated to the benchmark's maintenance and evolution is desirable.
\section{Ecoscape Suite}
\label{sec:Ecoscape}

\eval{} is designed as a benchmark tool to meet the requirements presented in the previous chapter. Deployed on a Kubernetes cluster, \eval{} offers a controlled and reproducible experiment environment, eliminating the need for a physical edge testbed. Hence, the inherent challenges of edge computing are simulated by \eval{}.

\eval{}'s operation is driven by a declarative configuration approach. Developers define their desired simulation scenario by providing a configuration file in JSON format. This configuration file serves as the blueprint for the test environment. These JSON configurations are then translated into Kubernetes manifests, which are subsequently deployed in the Kubernetes cluster. The configuration file is logically divided into four key sections: system, infrastructure, data, and chaos definition. The system's definition encompasses the distributed system's configuration, including the initial CPU and memory resource limits, the number of replicas, and the system-specific parameters passed as environment variables. The infrastructure definition allows users to specify the characteristics of the network, including latency and bandwidth limitations between nodes. The data definition describes the data generation process, specifying parameters such as the sending rate and location of the generated load. Finally, the chaos definition specifies the artificially injected infrastructure faults.

\eval{} provides a comprehensive score to evaluate the operator's performance within the simulated environment. The benchmark run is subdivided into four phases: the warm-up phase, the evaluation phase, the chaos phase, and a tear-down phase.
The warm-up phase accommodates the initial setup and allows the system to start entirely. The configurable warm-up period can be defined before the actual evaluation phase begins. Each SLO is continuously monitored from the evaluation phase. During the evaluation phase, chaos begins. Here, the faults are injected, and the remediator is expected to take action. The tear-down phase stops the system.   

The monitored SLOs are assigned a predefined weight that reflects their relative importance.
The final score provided by \eval{} is calculated as stated in \Cref{eq:slo_violation_avg}, where a higher metric value signifies a violation of the SLO. Consequently, SLOs must be formulated so that SLI values larger than the threshold indicate an SLO violation.
Within the equation, $v$ indicates the metric value at a given time $t$, while $\tau$ expresses the SLO threshold.
The score is inspired by the scaling performance metric~\cite{Straesser2023}, which is a weighted sum of SLO violations and the relative resource provisioning cost. In our approach, resource provisioning is also modelled as an SLO.

\begin{equation}
    \label{eq:slo_violation_avg}
    \hat{V}_{SLO} = \frac{1}{|T|} \sum_{t=1}^{T} 
    \begin{cases}
        1 - \frac{\tau}{v(t)}, & \text{for } v(t) > \tau \\
        0, & \text{otherwise}
    \end{cases}
\end{equation}

Hence, the SLO violation score is normalized from $0$ to $1$, where an increased violation score indicates a degraded performance outcome. When multiple SLOs are simultaneously active, their respective violation scores are aggregated using a weighted summation, as depicted in \Cref{eq:slo_violation_total}.

\begin{equation}
    \label{eq:slo_violation_total}    
    V_{total} = \sum_{i=1}^{|SLO|} w_i \cdot \hat{V}_{SLO_i}
\end{equation}

To implement the simulated constraints, \eval{} leverages the chaos engineering tool \textit{Chaos Mesh}\footnote{\href{https://chaos-mesh.org}{https://chaos-mesh.org}}. Kubernetes deployments within the simulated environment are annotated with location labels. Chaos Mesh interprets these labels to introduce the desired network properties or stressor scenarios. For comprehensive monitoring of system performance, \eval{} integrates \textit{Prometheus}\footnote{\href{https://prometheus.io}{https://prometheus.io}}. This allows a widespread interface that custom metrics can extend. Furthermore, energy consumption data are collected and made available through \textit{Kepler}~\cite{Amaral2024}, which seamlessly integrates with the Prometheus interface, providing a unified view of system behavior and energy usage.

\section{Case Study: Edge ML Inference}
\label{sec:Evaluation}

\begin{figure}[htbp] 
    \centering
    \includegraphics[width=0.5\textwidth]{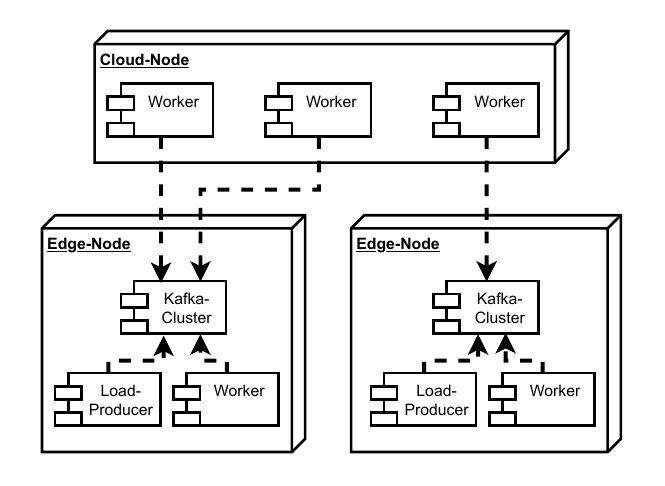}
    \caption{Deployment of the object recognition service on a cloud node and two edge nodes. Each edge node runs a Kafka cluster that stores local data from the load producer. Workers can be placed on any node. Each edge node serves one worker, and the cloud node serves three. }
    \label{fig:deployment}
\end{figure}

The suitability of \eval{} to evaluate Kubernetes remediation strategies is demonstrated by a case study. The case study focuses on an ML object recognition task based on the edge.
The object recognition service ingests incoming messages from a message broker, performs object recognition, and writes the results back to the originating message broker. The ResNet model trained on the ImageNet~\cite{Deng2009} dataset performs the object recognition and configures to variants of $50$, $101$, or $152$ hidden layers. The number of hidden layers can significantly affect the accuracy and speed of the classification task. Apache Kafka clusters are utilized as message brokers.
Our setup emulates a cloud node and two proximal edge nodes. 
Two Kafka clusters are deployed on the edge nodes, with load producers in the corresponding edge zones. Messages are sent to the proximal Kafka cluster, while the cloud node can consume messages from all clusters. The setup of the computing infrastructure is presented in \Cref{fig:deployment}.

The performance of the remediation strategies is evaluated based on three key SLOs: (1) accuracy, defined as the ratio of correctly classified images to the total number of classified images; (2) latency, measured as event-time latency \cite{Karimov2018}, including both queueing and processing times; and (3) energy consumption, quantified using the Kepler tool.

A functional prototype has been implemented and is publicly accessible on GitHub\footnote{\href{https://github.com/cau-se/Ecoscape}{https://github.com/cau-se/Ecoscape}} to evaluate the proposed \eval{} concept based on the described edge object recognition service.
While executing these chaos scenarios, we hypothesized that the defined SLOs would be violated. The specific SLOs under consideration for this evaluation were as follows:
(\textit{Processing Latency}) The end-to-end processing latency for the object recognition task must remain below $2.5\; seconds$.
(\textit{Object Recognition Accuracy}) The accuracy of the object recognition task, measured as the percentage of correctly classified objects, must exceed $75\; \%$.
(\textit{Energy Consumption}) The total energy consumption per object recognition task should not exceed $120\; Joules$. To emphasize the significance of real-time responses the latency SLO is weighted with 50\% while the energy consumption and the accuracy SLO are each weighted with 25\%. 
To simulate potential real-world challenges, two distinct fault scenarios were artificially introduced:
(\textit{Increased Network Latency}) Network latency was simulated by increasing message delay between nodes from an initial baseline of $50\; ms$ to $500\; ms$.
(\textit{CPU Stress on Edge Nodes}) CPU stress was artificially induced by concurrently running ten threads using the \Verb|stress-ng| utility. While executing these stress scenarios, we hypothesized that the defined SLOs, particularly those of latency, would be violated. Consequently, we anticipated that the remediator would initiate specific actions to restore SLO compliance. These expected remediation actions included:
(\textit{Model Depth Reduction}) Decreasing the complexity (depth) of the object recognition model deployed on the edge nodes to enhance their processing throughput and reduce latency.
(\textit{Workload Rescheduling}) Migrating the image classification workloads from the stressed edge nodes to available cloud nodes alleviating the computational burden on the edge infrastructure. 

\begin{figure}[htbp!]
    \centering
    \includegraphics[width=0.45\textwidth]{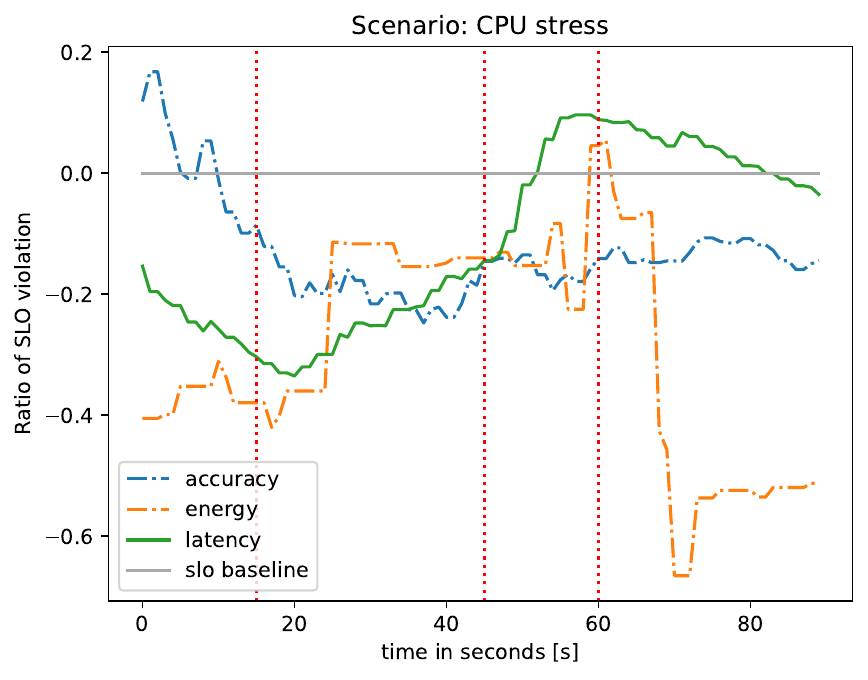}
    \caption{Ratio of the SLIs compared to the SLOs in the scenario of CPU stress on the edge nodes. After $15\; seconds$, chaos is injected into the system, violating the latency SLO. Between $45\; and\ 60\; seconds$, the system is reconfigured, approaching a non-violating state for the latency SLO. The total SLO violation score has a value of 0.011}
    \label{fig:cpu_stress}
\end{figure}

\begin{figure}[htbp!]
    \centering
    \includegraphics[width=0.45\textwidth]{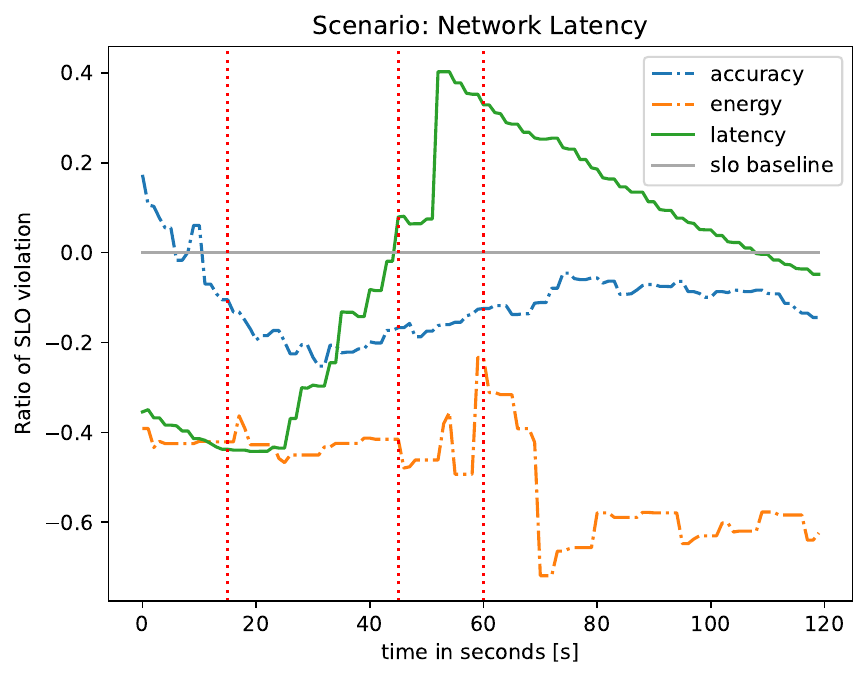}
    \caption{Ratio of the SLIs compared to the SLOs in the scenario of network latency between edge and cloud nodes. After $15\; seconds$, chaos is injected into the system, violating the latency SLO. Between $45\; and\ 60\; seconds$, the system is reconfigured, approaching a non-violating state for the latency SLO. The total SLO violation score has a value of 0.042}
    \label{fig:network_latency}
\end{figure}

The experimental evaluation was conducted on a Kubernetes cluster deployed within Kiel University's infrastructure. The initial system configuration comprised two edge nodes, each equipped with two available CPU cores and two cloud nodes with four available CPU cores. The SLIs during the benchmarks runs are published as a replication package \cite{Reiter25}.
\Cref{fig:cpu_stress,fig:network_latency} visualise the \eval{} benchmark runs. The first vertical red line signifies the transition from the warm-up phase to the evaluation phase. The predefined faults are injected into the environment when entering the evaluation phase. After 30 seconds, the remediator introduces a state change by acting to rectify the SLO violations. The second vertical red line illustrates this. The remediator finishes the reconfiguration process after 15 seconds as indicated by the third red line.

\Cref{fig:cpu_stress} illustrates an experiment in which CPU stress is added to the available edge nodes. As a result, latency increases and accuracy decreases as less CPU time is available for the ML inference task. Furthermore, energy consumption increases as a result of the increased CPU load. The remediator introduces a rescheduling action, migrating the ML inference to a cloud node. This introduces a transitional state, where neither the edge node nor the cloud node accepts any images to be inferred, as they either start up or shut down. The transitional state temporarily increases latency and energy consumption as the edge and cloud nodes run simultaneously. After the transitional state, latency steadily decreases, as images from the Kafka topic are consumed and inferred by the resourceful cloud node. 

\Cref{fig:network_latency} illustrates an experiment in which network latency is added to the available cloud nodes, making the edge nodes favourable. As a result, latency increases while accuracy decreases as images are still inferred on the cloud nodes, significantly increasing network latency. Energy consumption is unaffected as the edge nodes are fully utilised. The remediator introduces a model depth reduction task, reducing the hidden layers from $152$ to $50$. As previously mentioned, a transitional state is introduced where the nodes with the new model are starting up.

\section{Discussion}
\label{sec:Discussion}

The presented evaluation results demonstrate the proposed benchmark's potential to assess the suitability of various actions as responses to faults within edge computing environments. The correlation between the calculated scores and the trends observed in the graphical representations suggests a degree of internal consistency within the framework. Furthermore, the benchmark design allows for the specification of crucial experimental parameters, including experiment duration, system load, and variations in system configurations, thus satisfying essential requirements\cite{Hasselbring2021} of a benchmark.

Nevertheless, it is worth noting that the current evaluation did not involve an autonomous remediator. The absence of a complete remediator agent might preclude the observation of additional complexities and emergent behaviors. This initial study did not fully explore factors such as the remediator's decision-making logic, interaction with the Kubernetes API, and potential cascading effects on other system components. 
The feasibility of not utilizing a real edge computing testbed needs further consideration. CPU and memory requests and limits within Kubernetes assure computing resource availability for individual workloads. This suggests that the fundamental principles governing resource allocation and management remain applicable, even with a simplified representation of the underlying hardware. However, arguments against this abstraction highlight the inherent heterogeneity of edge devices and the potential for significant differences in CPU architectures, memory characteristics, and the availability of specialized hardware. 

Finally, it is pertinent to note that the proposed benchmark is currently in its early stages of development. The absence of an independent community actively contributing to its evolution and the lack of widespread adoption indicate that the tool's maturity is not yet fully established. Developing a collaborative community and further validation through diverse use cases and real-world deployments will enhance the benchmark's robustness, generalizability, and overall impact on the field.
\section{Conclusion}
\label{sec:Conclusion}

This paper addressed the challenge of establishing comprehensive benchmarks for evaluating remediation strategies to mitigate faults within edge computing environments. We formulated the problem by identifying the need for standardized methodologies to assess the efficacy of automated fault management in distributed edge infrastructures. To this end, we proposed the \textit{SLO violation score} that captures the performance of remediators to respond to performance degradation due to hardware faults.
To illustrate the practical application of our proposed framework \eval{}, we established a representative use case centred on object recognition in an edge computing setting. We investigated the dynamic behavior of key SLIs, explicitly processing latency, energy consumption, and ML inference accuracy under induced fault conditions. This evaluation provides empirical insights into the ability of a remediator reacting to fault scenarios to still comply with critical SLOs.

Building upon the findings presented herein, future research efforts will focus on further developing and refining the proposed benchmark. This includes fostering a community-driven initiative to establish a cornerstone for systematically evaluating remediators operating under under multiple SLOs. Furthermore, we aim to expand the scope of our benchmarking suites to encompass a broader spectrum of edge computing use cases and fault scenarios, thereby providing a more extensive and robust evaluation platform for automated remediation strategies.

\section*{Acknowledgement}
This work is supported by DIREC – Digital Research Centre Denmark and received funding from the Deutsche Forschungsgemeinschaft (DFG), grant 496119880

\bibliography{bibliography}

\begin{thebibliography}{10}
\providecommand{\url}[1]{#1}
\csname url@samestyle\endcsname
\providecommand{\newblock}{\relax}
\providecommand{\bibinfo}[2]{#2}
\providecommand{\BIBentrySTDinterwordspacing}{\spaceskip=0pt\relax}
\providecommand{\BIBentryALTinterwordstretchfactor}{4}
\providecommand{\BIBentryALTinterwordspacing}{\spaceskip=\fontdimen2\font plus
\BIBentryALTinterwordstretchfactor\fontdimen3\font minus \fontdimen4\font\relax}
\providecommand{\BIBforeignlanguage}[2]{{%
\expandafter\ifx\csname l@#1\endcsname\relax
\typeout{** WARNING: IEEEtran.bst: No hyphenation pattern has been}%
\typeout{** loaded for the language `#1'. Using the pattern for}%
\typeout{** the default language instead.}%
\else
\language=\csname l@#1\endcsname
\fi
#2}}
\providecommand{\BIBdecl}{\relax}
\BIBdecl

\bibitem{Chen2020}
\BIBentryALTinterwordspacing
S.~Chen, X.~Zhu, H.~Zhang, C.~Zhao, G.~Yang, and K.~Wang, ``Efficient privacy preserving data collection and computation offloading for fog-assisted iot,'' \emph{IEEE Transactions on Sustainable Computing}, vol.~5, no.~4, p. 526–540, Oct. 2020. [Online]. Available: \url{http://dx.doi.org/10.1109/TSUSC.2020.2968589}
\BIBentrySTDinterwordspacing

\bibitem{Meuser2024}
T.~Meuser, L.~Loven, M.~Bhuyan, S.~G. Patil, S.~Dustdar, A.~Aral, S.~Bayhan, C.~Becker, E.~D. Lara, A.~Y. Ding, J.~Edinger, J.~Gross, N.~Mohan, A.~D. Pimentel, E.~Riviere, H.~Schulzrinne, P.~Simoens, G.~Solmaz, and M.~Welzl, ``Revisiting edge ai: Opportunities and challenges,'' \emph{IEEE Internet Computing}, vol.~28, pp. 49--59, 2024.

\bibitem{Bian2022}
\BIBentryALTinterwordspacing
J.~Bian, A.~A. Arafat, H.~Xiong, J.~Li, L.~Li, H.~Chen, J.~Wang, D.~Dou, and Z.~Guo, ``Machine learning in real-time internet of things (iot) systems: A survey,'' \emph{IEEE Internet of Things Journal}, vol.~9, no.~11, p. 8364–8386, Jun. 2022. [Online]. Available: \url{http://dx.doi.org/10.1109/JIOT.2022.3161050}
\BIBentrySTDinterwordspacing

\bibitem{Molchanov2016}
\BIBentryALTinterwordspacing
P.~Molchanov, S.~Tyree, T.~Karras, T.~Aila, and J.~Kautz, ``Pruning convolutional neural networks for resource efficient inference,'' 2016. [Online]. Available: \url{https://arxiv.org/abs/1611.06440}
\BIBentrySTDinterwordspacing

\bibitem{Jain2018}
\BIBentryALTinterwordspacing
S.~Jain, S.~Venkataramani, V.~Srinivasan, J.~Choi, P.~Chuang, and L.~Chang, ``Compensated-dnn: energy efficient low-precision deep neural networks by compensating quantization errors,'' in \emph{Proceedings of the 55th Annual Design Automation Conference}, ser. DAC ’18.\hskip 1em plus 0.5em minus 0.4em\relax ACM, Jun. 2018, p. 1–6. [Online]. Available: \url{http://dx.doi.org/10.1145/3195970.3196012}
\BIBentrySTDinterwordspacing

\bibitem{Panda16}
P.~Panda, A.~Sengupta, and K.~Roy, ``Conditional deep learning for energy-efficient and enhanced pattern recognition,'' in \emph{Proceedings of the 2016 Conference on Design, Automation \& Test in Europe}, ser. DATE '16.\hskip 1em plus 0.5em minus 0.4em\relax San Jose, CA, USA: EDA Consortium, 2016, p. 475–480.

\bibitem{Pourreza2023}
\BIBentryALTinterwordspacing
M.~Pourreza and P.~Narasimhan, ``A survey of faults and fault-injection techniques in edge computing systems,'' in \emph{2023 IEEE International Conference on Edge Computing and Communications (EDGE)}.\hskip 1em plus 0.5em minus 0.4em\relax IEEE, Jul. 2023, p. 63–71. [Online]. Available: \url{http://dx.doi.org/10.1109/EDGE60047.2023.00021}
\BIBentrySTDinterwordspacing

\bibitem{Tran2022}
\BIBentryALTinterwordspacing
M.-N. Tran, D.-D. Vu, and Y.~Kim, ``A survey of autoscaling in kubernetes,'' in \emph{2022 Thirteenth International Conference on Ubiquitous and Future Networks (ICUFN)}.\hskip 1em plus 0.5em minus 0.4em\relax IEEE, Jul. 2022. [Online]. Available: \url{http://dx.doi.org/10.1109/ICUFN55119.2022.9829572}
\BIBentrySTDinterwordspacing

\bibitem{Senjab2023}
\BIBentryALTinterwordspacing
K.~Senjab, S.~Abbas, N.~Ahmed, and A.~u.~R. Khan, ``A survey of kubernetes scheduling algorithms,'' \emph{Journal of Cloud Computing}, vol.~12, no.~1, Jun. 2023. [Online]. Available: \url{http://dx.doi.org/10.1186/s13677-023-00471-1}
\BIBentrySTDinterwordspacing

\bibitem{Rejiba2022}
\BIBentryALTinterwordspacing
Z.~Rejiba and J.~Chamanara, ``Custom scheduling in kubernetes: A survey on common problems and solution approaches,'' \emph{ACM Computing Surveys}, vol.~55, no.~7, p. 1–37, Dec. 2022. [Online]. Available: \url{http://dx.doi.org/10.1145/3544788}
\BIBentrySTDinterwordspacing

\bibitem{Hamid24}
\BIBentryALTinterwordspacing
A.~R. Hamid, H.~Reiter, M.~B. Kj{\ae}rgaard, and W.~Hasselbring, ``Investigating quality attributes of machine learning inference on the edge-cloud continuum,'' \emph{Softwaretechnik-Trends}, vol.~45, no.~1, November 2024, proceeddings 15th Symposium on Software Performance. [Online]. Available: \url{https://oceanrep.geomar.de/id/eprint/62115/}
\BIBentrySTDinterwordspacing

\bibitem{Nikolaidis2021}
\BIBentryALTinterwordspacing
F.~Nikolaidis, A.~Chazapis, M.~Marazakis, and A.~Bilas, ``Frisbee: A suite for benchmarking systems recovery,'' in \emph{Proceedings of the 1st Workshop on High Availability and Observability of Cloud Systems}, ser. EuroSys ’21.\hskip 1em plus 0.5em minus 0.4em\relax ACM, Apr. 2021, p. 18–24. [Online]. Available: \url{http://dx.doi.org/10.1145/3447851.3458738}
\BIBentrySTDinterwordspacing

\bibitem{Kalka2024}
\BIBentryALTinterwordspacing
W.~Kalka and T.~Szydlo, \emph{$\mu$Chaos: Moving Chaos Engineering to IoT Devices}.\hskip 1em plus 0.5em minus 0.4em\relax Springer Nature Switzerland, 2024, p. 239–254. [Online]. Available: \url{http://dx.doi.org/10.1007/978-3-031-63783-4_18}
\BIBentrySTDinterwordspacing

\bibitem{Sonmez2017}
\BIBentryALTinterwordspacing
C.~Sonmez, A.~Ozgovde, and C.~Ersoy, ``Edgecloudsim: An environment for performance evaluation of edge computing systems,'' in \emph{2017 Second International Conference on Fog and Mobile Edge Computing (FMEC)}.\hskip 1em plus 0.5em minus 0.4em\relax IEEE, May 2017. [Online]. Available: \url{http://dx.doi.org/10.1109/FMEC.2017.7946405}
\BIBentrySTDinterwordspacing

\bibitem{Moreschini2022}
\BIBentryALTinterwordspacing
S.~Moreschini, F.~Pecorelli, X.~Li, S.~Naz, D.~Hastbacka, and D.~Taibi, ``Cloud continuum: The definition,'' \emph{IEEE Access}, vol.~10, p. 131876–131886, 2022. [Online]. Available: \url{http://dx.doi.org/10.1109/ACCESS.2022.3229185}
\BIBentrySTDinterwordspacing

\bibitem{McChesney2019}
\BIBentryALTinterwordspacing
J.~McChesney, N.~Wang, A.~Tanwer, E.~de~Lara, and B.~Varghese, ``Defog: fog computing benchmarks,'' in \emph{Proceedings of the 4th ACM/IEEE Symposium on Edge Computing}, ser. SEC ’19.\hskip 1em plus 0.5em minus 0.4em\relax ACM, Nov. 2019, p. 47–58. [Online]. Available: \url{http://dx.doi.org/10.1145/3318216.3363299}
\BIBentrySTDinterwordspacing

\bibitem{Wan2022}
\BIBentryALTinterwordspacing
Z.~Wan, Z.~Zhang, R.~Yin, and G.~Yu, ``Kfiml: Kubernetes-based fog computing iot platform for online machine learning,'' \emph{IEEE Internet of Things Journal}, vol.~9, no.~19, p. 19463–19476, Oct. 2022. [Online]. Available: \url{http://dx.doi.org/10.1109/JIOT.2022.3168085}
\BIBentrySTDinterwordspacing

\bibitem{Santos2019}
\BIBentryALTinterwordspacing
J.~Santos, T.~Wauters, B.~Volckaert, and F.~De~Turck, ``Towards network-aware resource provisioning in kubernetes for fog computing applications,'' in \emph{2019 IEEE Conference on Network Softwarization (NetSoft)}.\hskip 1em plus 0.5em minus 0.4em\relax IEEE, Jun. 2019, p. 351–359. [Online]. Available: \url{http://dx.doi.org/10.1109/NETSOFT.2019.8806671}
\BIBentrySTDinterwordspacing

\bibitem{Eidenbenz2020}
\BIBentryALTinterwordspacing
R.~Eidenbenz, Y.-A. Pignolet, and A.~Ryser, ``Latency-aware industrial fog application orchestration with kubernetes,'' in \emph{2020 Fifth International Conference on Fog and Mobile Edge Computing (FMEC)}.\hskip 1em plus 0.5em minus 0.4em\relax IEEE, Apr. 2020, p. 164–171. [Online]. Available: \url{http://dx.doi.org/10.1109/FMEC49853.2020.9144934}
\BIBentrySTDinterwordspacing

\bibitem{Deng2016}
\BIBentryALTinterwordspacing
R.~Deng, R.~Lu, C.~Lai, T.~H. Luan, and H.~Liang, ``Optimal workload allocation in fog-cloud computing towards balanced delay and power consumption,'' \emph{IEEE Internet of Things Journal}, p. 1–1, 2016. [Online]. Available: \url{http://dx.doi.org/10.1109/JIOT.2016.2565516}
\BIBentrySTDinterwordspacing

\bibitem{Salfner2010}
\BIBentryALTinterwordspacing
F.~Salfner, M.~Lenk, and M.~Malek, ``A survey of online failure prediction methods,'' \emph{ACM Computing Surveys}, vol.~42, no.~3, p. 1–42, Mar. 2010. [Online]. Available: \url{http://dx.doi.org/10.1145/1670679.1670680}
\BIBentrySTDinterwordspacing

\bibitem{Owotogbe2024}
\BIBentryALTinterwordspacing
J.~Owotogbe, I.~Kumara, W.-J. V.~D. Heuvel, and D.~A. Tamburri, ``Chaos engineering: A multi-vocal literature review,'' 2024. [Online]. Available: \url{https://arxiv.org/abs/2412.01416}
\BIBentrySTDinterwordspacing

\bibitem{Qazi2024}
\BIBentryALTinterwordspacing
F.~Qazi, D.~Kwak, F.~G. Khan, F.~Ali, and S.~U. Khan, ``Service level agreement in cloud computing: Taxonomy, prospects, and challenges,'' \emph{Internet of Things}, vol.~25, p. 101126, Apr. 2024. [Online]. Available: \url{http://dx.doi.org/10.1016/j.iot.2024.101126}
\BIBentrySTDinterwordspacing

\bibitem{Hasselbring2021}
\BIBentryALTinterwordspacing
W.~Hasselbring, ``Benchmarking as empirical standard in software engineering research,'' in \emph{Evaluation and Assessment in Software Engineering}, ser. EASE 2021.\hskip 1em plus 0.5em minus 0.4em\relax ACM, Jun. 2021, p. 365–372. [Online]. Available: \url{http://dx.doi.org/10.1145/3463274.3463361}
\BIBentrySTDinterwordspacing

\bibitem{Straesser2023}
\BIBentryALTinterwordspacing
M.~Straesser, S.~Eismann, J.~von Kistowski, A.~Bauer, and S.~Kounev, ``Autoscaler evaluation and configuration: A practitioner’s guideline,'' in \emph{Proceedings of the 2023 ACM/SPEC International Conference on Performance Engineering}, ser. ICPE ’23.\hskip 1em plus 0.5em minus 0.4em\relax ACM, Apr. 2023, p. 31–41. [Online]. Available: \url{http://dx.doi.org/10.1145/3578244.3583721}
\BIBentrySTDinterwordspacing

\bibitem{Amaral2024}
M.~Amaral, H.~Chen, T.~Chiba, R.~Nakazawa, S.~Choochotkaew, E.~K. Lee, and T.~Eilam, ``{Process-Based Efficient Power Level Exporter},'' \emph{IEEE International Conference on Cloud Computing, CLOUD}, pp. 456--467, 2024.

\bibitem{Deng2009}
\BIBentryALTinterwordspacing
J.~Deng, W.~Dong, R.~Socher, L.-J. Li, K.~Li, and L.~Fei-Fei, ``Imagenet: A large-scale hierarchical image database,'' in \emph{2009 IEEE Conference on Computer Vision and Pattern Recognition}.\hskip 1em plus 0.5em minus 0.4em\relax IEEE, Jun. 2009. [Online]. Available: \url{http://dx.doi.org/10.1109/CVPR.2009.5206848}
\BIBentrySTDinterwordspacing

\bibitem{Karimov2018}
\BIBentryALTinterwordspacing
J.~Karimov, T.~Rabl, A.~Katsifodimos, R.~Samarev, H.~Heiskanen, and V.~Markl, ``Benchmarking distributed stream data processing systems,'' in \emph{2018 IEEE 34th International Conference on Data Engineering (ICDE)}.\hskip 1em plus 0.5em minus 0.4em\relax IEEE, Apr. 2018. [Online]. Available: \url{http://dx.doi.org/10.1109/ICDE.2018.00169}
\BIBentrySTDinterwordspacing

\bibitem{Reiter25}
\BIBentryALTinterwordspacing
H.~Reiter and A.~R. Hamid, ``Ecoscape slis,'' 2025. [Online]. Available: \url{https://zenodo.org/doi/10.5281/zenodo.15170211}
\BIBentrySTDinterwordspacing

\end{thebibliography}
\bibliographystyle{IEEEtran}

\end{document}